\documentclass[conference]{IEEEtran}
\IEEEoverridecommandlockouts

\usepackage{cite}
\usepackage{amsmath,amssymb,amsfonts}
\usepackage{algorithmic}
\usepackage{graphicx}
\usepackage{textcomp}
\usepackage{xcolor}
\usepackage{csvsimple}
\usepackage{rotating}
\usepackage{subfig}
\usepackage{float}
\usepackage{url}
\usepackage[skip=8pt,font=scriptsize]{caption}

\usepackage{listings}
\usepackage{color}

\definecolor{dkgreen}{rgb}{0,0.6,0}
\definecolor{gray}{rgb}{0.5,0.5,0.5}
\definecolor{mauve}{rgb}{0.58,0,0.82}

\usepackage{wrapfig}
\usepackage{lscape}
\usepackage{rotating}
\usepackage{epstopdf}

\def\BibTeX{{\rm B\kern-.05em{\sc i\kern-.025em b}\kern-.08em
		T\kern-.1667em\lower.7ex\hbox{E}\kern-.125em}}
\graphicspath{ {fig/} }

\begin{document}
	
	\title{Efficient Generation of Mandelbrot Set using Message Passing Interface}
	
	\author{
		\IEEEauthorblockN{Bhanuka Manesha Samarasekara Vitharana Gamage\IEEEauthorrefmark{1},
		Vishnu Monn Baskaran\IEEEauthorrefmark{2}}
		\IEEEauthorblockA{School of Information Technology,\\Monash University Malaysia,\\ Jalan Lagoon Selatan, 47500 Bandar Sunway,\\ Selangor Darul Ehsan, Malaysia.\\
		Email: \IEEEauthorrefmark{1}bsam0002@student.monash.edu,
		\IEEEauthorrefmark{2}vishnu.monn@monash.edu}
	}
		
	\maketitle
	
	\begin{abstract}
		With the increasing need for safer and reliable systems, Mandelbrot Set's use in the encryption world is evident to everyone. This document aims to provide an efficient method to generate this set using data parallelism. First Bernstein's conditions are used to ensure that the Data is parallelizable when generating the Mandelbrot Set. Then Amdhal's Law is used to calculate the theoretical speed up, to be used to compare three partition schemes. The three partition schemes discussed in this document are the Naïve Row Segmentation, the First Come First Served Row Segmentation and the Alternating Row Segmentation. The Message Parsing Interface (MPI) library in C is used for all of the communication. After testing all the implementation on MonARCH, the results demonstrate that the Naïve Row Segmentation approach did not perform as par. But the Alternating Row Segmentation approach performs better when the number of tasks are $< 16$, where as the First Come First Served approach performs better when the number of tasks is $\ge 16$. 
	\end{abstract}

	
	\section{Introduction}

	The Mandelbrot set, named after Benoit Mandelbrot is the set of points c in the complex plane which produces a bounded sequence, with the application of equation \ref{MandelbrotEq} repeatedly to the point $z=0$ \cite{Mandelbrotset}. Aside from the inherent beauty of the Mandelbrot Set, some of the mysteries of this set is still unknown to Mathematicians \cite{lamb_2017}. So having a way to generate this set efficiently can increase the research in this area and improve our understanding of this set. 
	\begin{equation}
		f(z)=z^{2} + c \label{MandelbrotEq}
	\end{equation}
	
	In order to generate the Mandelbrot set for a $m \times n$ image $I$, we need to execute equation \ref{MandelbrotEq} on each pixel of the image. The complex part of the equation is determined using the $m$ and $n$ values of the image. Then the Iteration at which either the magnitude ($z^2$) exceeding the escape radius or the \textit{iterationMax}, is recorded. Both the \textit{iterationMax} and Escape Radius are pre-determined and can affect the time taken to generate the set. The iteration is then used to determine the pixel color in the image $I$. Therefore each pixel of the image can be represented using equation \ref{MandelPixel}. 
	
\begin{figure}[h]
	\centering
	\includegraphics[width=1.9in]{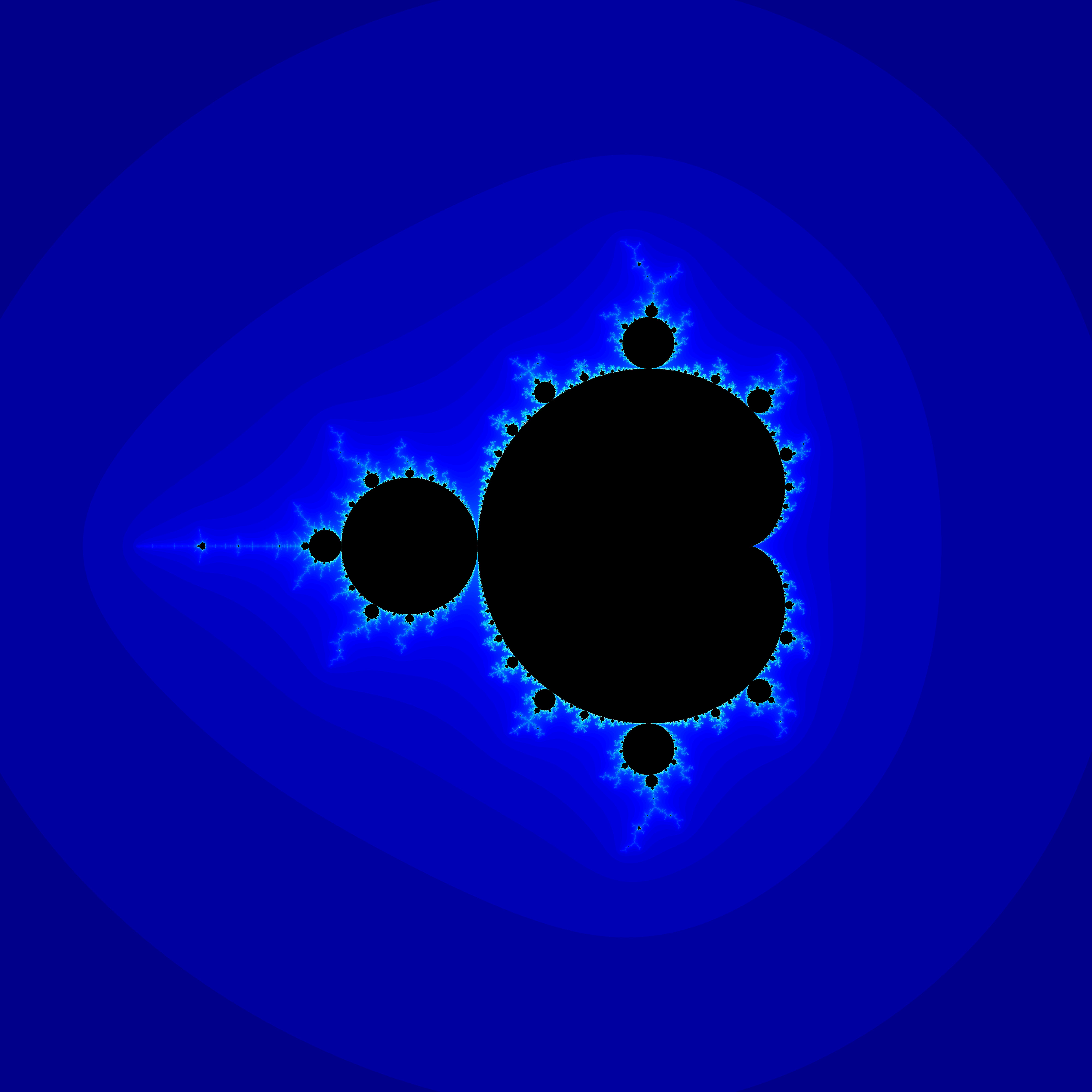}
	\caption{Mandelbrot Set generated using Boolean Escape Time algorithm\cite{rosettacode}}
	\label{Mandel}
\end{figure}

	This method is known as Mandelbrot generation with Boolean escape time. Figure \ref{Mandel} shows the image of a Mandelbrot Set generated using this method.
			\begin{equation}
	I_{i,j} = \text{\emph{Iteration}}_{i,j}  \label{MandelPixel}
	\end{equation}
	\begin{center}
		where $i = 1...m, j = 1...n$
	\end{center}

	There are many uses of the Mandelbrot Set in the field of encryption such as Image Encryption\cite{Iencryption}, Key Exchange Protocols \cite{alia2007new} and Key Encryption\cite{agarwal2017symmetric}. With the increasing need for cyber-security in the modern era, safer and reliable encryption techniques will play a major role in the future.
 	
	This report aims to provide an efficient partition scheme to generate the Mandelbrot Set. For this, three partition schemes are compared and contrasted against each other and then ranked based on their performance. 
	First, Bernstein's conditions are used to determine whether the generation of Mandelbrot Set is data parallelizable \cite{bernstein1996analysis}. Then the theoretical speed up is calculated using Amdhal's law \cite{amdahl1967validity}, which is used as the objective for each partition scheme.  The workload is divided among $N$ number of tasks, in different segmentation configurations for each of the partition schemes. Then the actual speed up is obtained for all of the partition schemes with different task configurations. Using the theoretical speed up and the actual speed up, the percentage difference of each of the partition schemes are then obtained and used for the comparison. All of the inter process communication is done using the Message Parsing Library (MPI) in C \cite{openmpi}.

	\section{Preliminary Analysis}
	In this section, we examine whether the generation of the Mandelbrot Set is data parallelizable and if there is any advantage of doing it.
			\subsection{Data Parallelizability of the Generation of the Mandelbrot Set}
			
			In order to ensure that the process of generating the Mandelbrot Set is data parallelizable, we can use Bernstein’s conditions \cite{bernstein1996analysis}. Bernstein's conditions are a simple verification for deciding if operations and statements can work simultaneously without changing the program output and allow for data parallelism \cite{Feautrier2011}. According to Bernstein, if two tasks satisfy the following equations, then they are parallelizable.
			
			\begin{equation}
				I_1 \cap O_2 = \emptyset\label{anti independency}
			\end{equation}
			\begin{equation}
				I_2 \cap O_1 = \emptyset \label{flow independency}
			\end{equation}
			\begin{equation}
				O_1 \cap O_2 = \emptyset \label{output independency}
			\end{equation}
			
			$I_0$ and $I_1$ represents the inputs for the first and second tasks while  $O_0$ and $O_1$ represents the outputs from first and second tasks. Equation \ref{anti independency} also known as anti in-dependency, states that the input of the first task should not have any dependency with the output of the second task, while equation \ref{flow independency}, also known as flow in-dependency, states that the input of the second task should not have any dependency with the output of the first task. And finally, equation \ref{output independency} also known as output in-dependency, states that both the outputs of the two tasks should not be equal.
			
			Using equation \ref{MandelPixel}, we can derive the input and output equations for any two neighboring pixels in Image $I$ which will be computed by two tasks $P_{1}$ and $P_{2}$.
			
			\begin{equation}
			I_1 = \text{\emph{Iteration}}_{i,j} \label{I1}
			\end{equation}
			\begin{equation}
			O_1 = I_{i,j} \label{O1}
			\end{equation}
			\begin{equation}
			I_2= \text{\emph{Iteration}}_{i,j+1} \label{I2}
			\end{equation}
			\begin{equation}
			O_2= I_{i,j+1} \label{O2}
			\end{equation}
			
			Applying Bernstein conditions for equations \ref{I1}, \ref{O1}, \ref{I2} and \ref{O2}, we get,
			
			\begin{equation}
			\text{\emph{Iteration}}_{i,j}  \quad  \cap \quad I_{i,j+1} = \emptyset
			\end{equation}
			\begin{equation}
			\text{\emph{Iteration}}_{i,j+1} \quad  \cap \quad I_{i,j} = \emptyset 
			\end{equation}
			\begin{equation}
			I_{i,j}  \quad  \cap \quad  I_{i,j+1} = \emptyset 
			\end{equation}
			
			Since all the above conditions are satisfied, we can state that the generation of the Mandelbrot Set satisfies the Bernstein conditions, thereby the data can be parallelized  and the Mandelbrot Set can be generated in parallel. 
			
			Next we calculate the theoretical speed up to determine whether there is a benefit of running the code in parallel.

			\subsection{Theoretical Speed Up of the Generation of the  Mandelbrot Set}
			\label{Theo}
			To determine the theoretical speed up of the Mandelbrot Set Generation, we first determine the pararalizable portion of the serial implementation of the code. This can be achieved by calculating the total time for generating the Mandelbrot set ($t_p$) and the total time to execute the whole serial code ($t_s$). Then using equation \ref{rp} we can get the $r_p$ value.
			
			\begin{equation}
				r_p = \frac{t_p}{t_s} \label{rp}
			\end{equation}
			
			Then after calculating the $r_p$ value, we can use Amdahl's Law \cite{amdahl1967validity} to calculate the theoretical speed up to generate the Mandelbrot set. Amdahl stated that if $p$ is the number of tasks, $r_s$ is the time spent on the serial part of the code and $r_p$ is the amount of time spent on the part of the program that can be done in parallel \cite{gustafson1988reevaluating}, then the speed up can be stated as 
			
			\begin{equation}
				S_p = \frac{1}{r_s + \frac{r_p}{p}}
			\end{equation}
			
			Using Amdahl's law, we are able to generate Table \ref{theoratical} with the theoretical speed up values.
			
			\begin{table}[!h]\caption{Number of tasks vs Theoretical Speed up factor}
				\begin{center}
					\renewcommand{\arraystretch}{1.2}
					
					\begin{tabular}{|c|c|} 
						\hline
						\textbf{Number of Tasks} & \textbf{Speed up Factor} \\ \hline
						2                  & 1.99440                     \\
						4                  & 3.96659                     \\
						8                   & 7.84583                     \\
						16                 & 15.35350                       \\
						32                & 29.43820                      \\
						 $\infty$      & 356.22764                       \\ \hline
						 \multicolumn{2}{l}{\textit{More details in the Appendix - Figure \ref{app:amdal}}	 }
					\end{tabular}
				\label{theoratical}
				\end{center}
			\end{table}
			
			So having determined that the generation of the Mandelbrot Set is embarrassingly parallelizable and having calculated the theoretical speed up, next we look into the designs of the partition schemes for parallelizing data.
	
	\begin{figure*}[t!]
	\centering
	\includegraphics[width=7in]{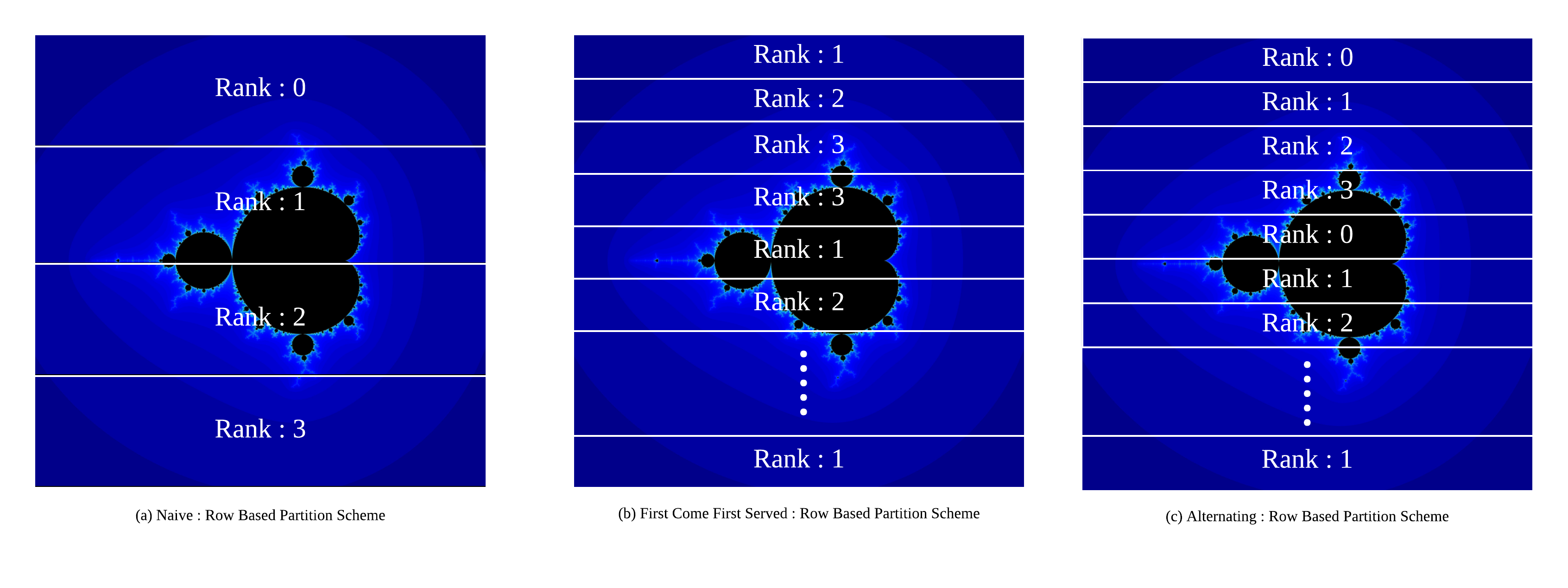} 
	\caption{Partition Schemes when $N=4$ }
	\label{Partition}
\end{figure*}
			
	\section{Design of the Partition Schemes}
	For the design of the partition schemes, we will be comparing three partition schemes. In all of these schemes, we will be writing the Mandelbrot Set to the file in only the master node. So each approach will send the data back to the master node, therefore all of them will be using a Master-Slave Architecture. Each partition scheme will be under a controlled environment, thus allowing us to compare the actual Mandelbrot Generation speed up.

			\subsection{Naive : Row Based Partition Scheme }\label{Naive}
			As shown in Figure \ref{Partition}, the final image with $iYmax$ rows is divided equally among N number of processors. So each process will pre-calculate its start and end positions and generate the Mandelbrot Set for that portion. Equation \ref{stp} , \ref{etp} and \ref{rem} are used to calculate the start ($S_r$) and end point ($E_r$)  for each processor with rank ($r$) out of $N$ tasks.
			
			\begin{equation}
			S_r = \frac{iYmax}{N} \times r
			\label{stp}
			\end{equation}
			
			\begin{equation}
			E_r = \frac{iYmax}{N} \times (r \times 1 ) - 1
			\label{etp}
			\end{equation}
			
			Using this partition scheme each process gets an equal amount of rows to work with, which can be written using the following equation.
			\begin{equation}
				\label{rowsPerTask}
				\textit{rows per task} = \frac{iYmax}{\textit{number of tasks}}
			\end{equation}
			
			In the case where the number of rows cannot be equally divided among the number of tasks ($N$), the last node (rank = $N$ - 1) will calculate the remaining set of rows. This will not effect the overall performance significantly since the top and bottom section of the Mandelbrot Set takes less time to process compared to the middle section. So the end point ($E_r$) for the last task can be represented by Equation \ref{rem}.
			
			\begin{equation}
			 E_r = E_r + (iYmax \mod N)
			\label{rem}
			\end{equation}
			
			After generating the $S_r$ and $E_r$ points,  all of the nodes will generate the $Cy$ and $Cx$ arrays. Then the master node will open the file and it will start generating the Mandelbrot Set for the rows allocated for it. The other nodes will also use the $S_r$ and $E_r$ points to generate their specific rows. Finally, all the slave nodes will send their generated values to the master node. The master node will then receive the chunks of rows and write it to the file. 
			
			\begin{equation}
			T_m = N - 1
			\label{naive_m}
			\end{equation}
			
\begin{figure}[ht]
			\centering
			\includegraphics[width=3in]{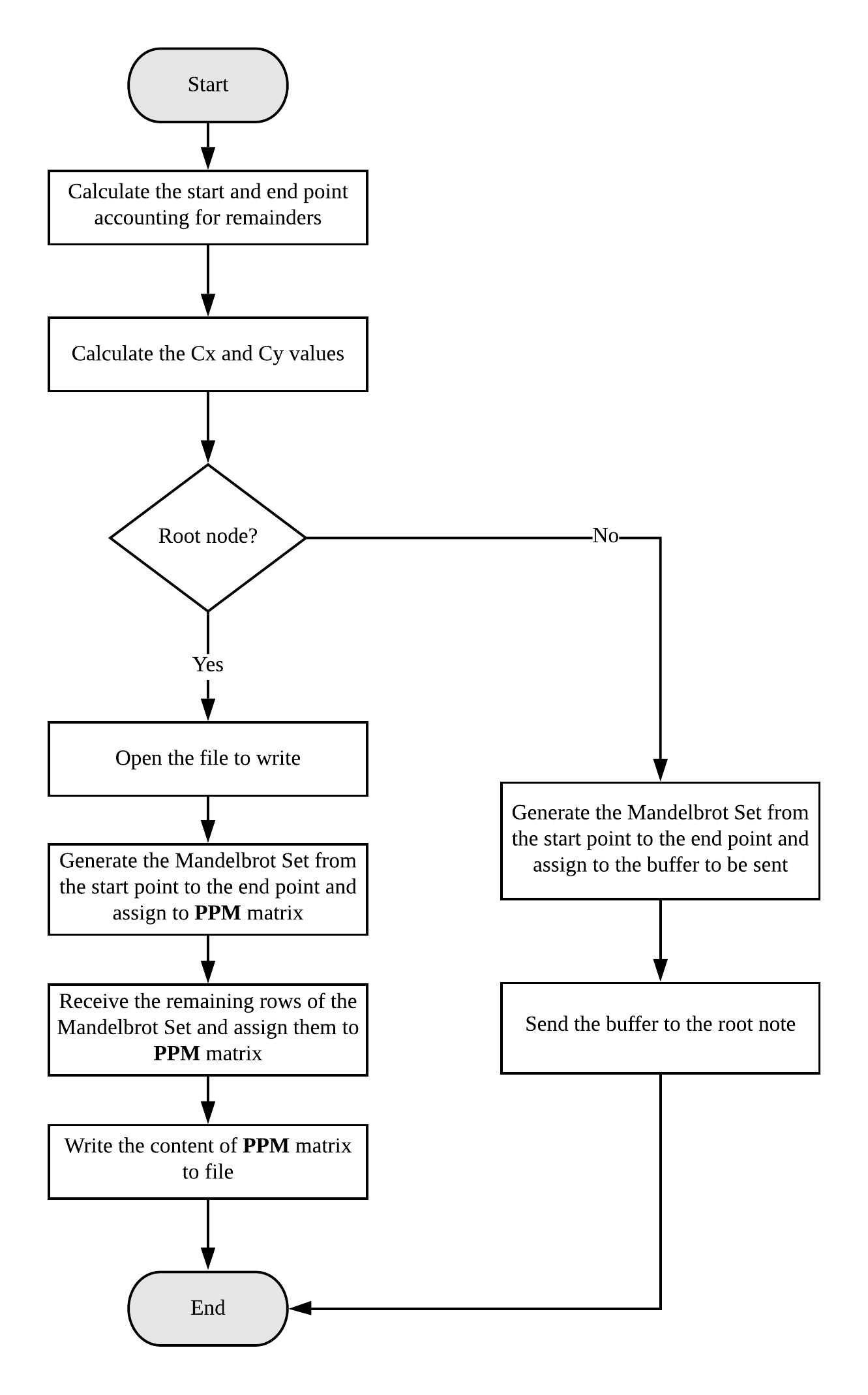}
			\caption{Technical Flowchart for Naïve Row Based Partition Scheme}
			\label{Naive_1}
			\end{figure}

			Equation \ref{naive_m} shows the ($T_M$) total number of messages passed between N number of tasks, when using the Naive Partition Scheme. The technical flowchart for the Naïve row based partition scheme is shown in Figure \ref{Naive_1}.

			\begin{figure*}[ht]
				\centering
				\includegraphics[width=7in]{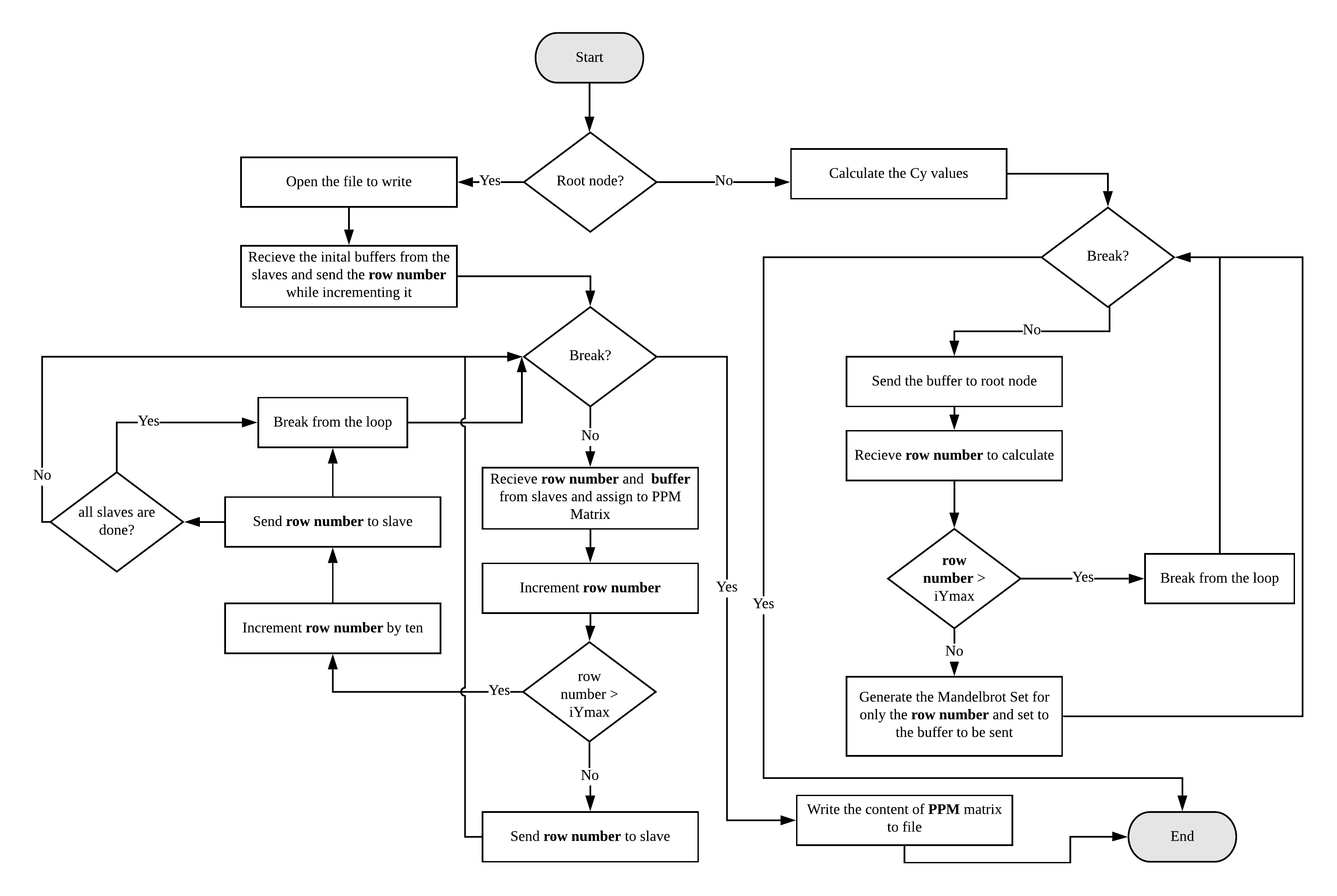}
				\caption{Technical Flowchart for First Come First Served Row Based Partition Scheme}
				\label{FCFS_i}
			\end{figure*}
			
			\subsection{First Come First Served : Row Based Partition Scheme}\label{FCFS}
			This partition scheme shares similarity with schedulers in operating systems, where the master node sends work, based on the availability. As shown in Figure \ref{Partition}, the order in which node performs which row is based on the availability of the node. If one node finishes the row allocated to it, then it will proceed to the next row. So the master node keeps track of the rows that are processed so far and it will send the next row to the next available task. The slave node will then receive the row number and calculate the Mandelbrot set for that row. This will be done until the master node sends a row number larger than $iYmax$. At that point the node has finished all of the work, so it will stop and exit the program. The master node as it receives will copy each row from the slaves and use the row number to decide which row it is and add it to the memory. Then after receiving all the rows, it will write the data from memory to file. For this partition scheme, the issue of having an odd task count or odd number of rows is not present, since its based on the availability of tasks. 
			Due to the nature of work allocation, this partition scheme is named as First Come First Served. The main thing to note is that the master node does not generate any rows. It will only be delegating work to the slave nodes. Another observation would be that this method has a high communication overhead as the master node has to send each row number to the slaves and then receive each row from the slaves.
			
			\begin{equation}
			T_m = iYmax \times 2
			\label{fcfs_m}
			\end{equation}
			
			Equation \ref{fcfs_m} shows the ($T_M$) total number of messages passed between N number of tasks, when using the First Come First Served Partition Scheme. So we can observe that the total number of tasks does not effect the number of messages when using this approach.The technical flowchart for the First Come First Served row based partition scheme is shown in Figure \ref{FCFS_i}.

			\subsection{Alternating : Row Based Partition Scheme}\label{Alternating}
			This partition scheme is similar to \ref{Naive}, where all the tasks divide the work among themselves initially. Figure \ref{Partition} shows that each task will perform the alternating rows, thus dividing the number of rows equally. In the case where the number of rows cannot be equally divided, the master node will then perform the remainder. Initially all the nodes calculate the $Cy$ and $Cx$ values. The master node then opens the file and starts calculating the rows, starting from the $0^{th}$ index until the end, with increments of the number of tasks. Similarly, each node starts from their respective rank and increments with the number of tasks. Similar to \ref{Naive}, each process gets an equal amount of rows to work with, which can be written using equation \ref{rowsPerTask}. Then all the slave nodes sends back the generated rows to the master node. Before receiving, if there is a remainder, the master node will increment one by one to calculate it. Similar to \ref{Naive}, the processing time for the remainder is insignificant compared to the rest. After receiving all the rows, the master node will loop though the $0^{th}$ index till $iYmax$ and reconstruct the image collecting alternate rows from the received arrays. Then it will write all the data into the file.
					
			\begin{equation}
			T_m =  N - 1
			\label{al_m}
			\end{equation}
			
\begin{figure}[ht]
				\centering
				\includegraphics[width=3in]{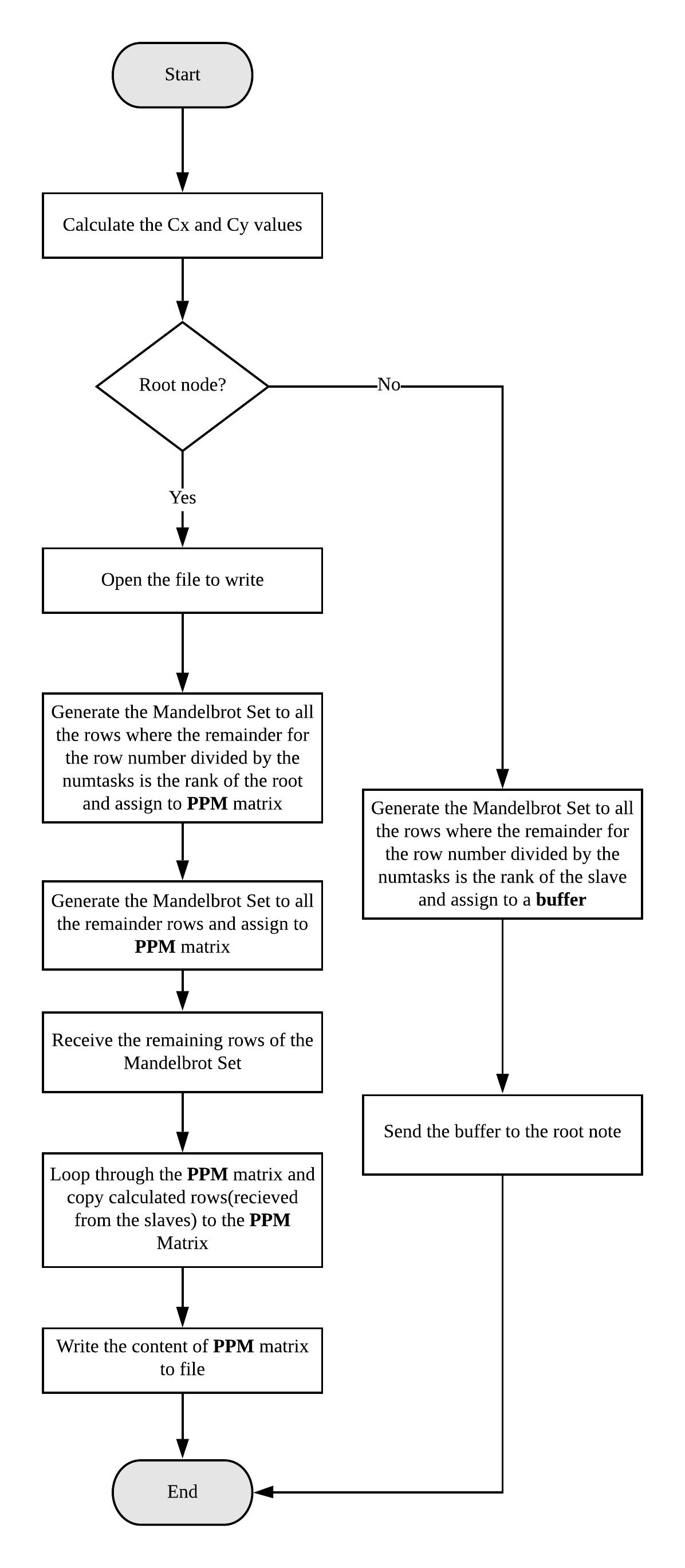}
				\caption{Technical Flowchart for Alternative Row Based Partition Scheme}
				\label{Alte}
			\end{figure}

			Equation \ref{al_m} shows the ($T_M$) total number of messages passed between N number of tasks, when using the Alternating Partition Scheme. The technical flowchart for the Alternative row based partition scheme is shown in Figure \ref{Alte}.

	\section{Methodology}
	 \label{TEST}
	In order to calculate the run times for the three partition schemes, multiple tests were done. For each test, we set the size of the image to 8000 $\times$ 8000 with 2000 iterations and the Escape Radius is set to 400. An assumption used for the test cases was that the task count will always be a multiple of two, but all of these partition schemes will work with any other task count, as all of them have a method to handle remainders. Each partition scheme was executed 5 times, with 2, 4, 8, 16 and 32 tasks. \newline
	All of the tests were done on the MonARCH (Monash Advanced Research Computing Hybrid) HPC/HTC cluster \cite{mon}. The problem with using a cluster is that the scheduler decides which CPU to use based on some predefined rules such as availability. So for our tests to be equal, a constraint was added in order to limit the tests to only one CPU type. All the tests were done on Intel® Xeon® Gold 6150 Processors with 36 logical cores. But since hyper threading was turned off, each processor had only 18 cores. So in order to run the 32 task test, two 16 core processors were used. Listing \ref{16job} and \ref{32job} of the appendix, shows the job file used to run the parallel code with 16 and 32 cores respectively. The RAM was set to 32GB for all the tests. \newline
	For each test, we calculated the time taken for generating the Mandelbrot Set and also the total time taken, including the time taken to write data to the file. All the results (Figure \ref{app:n}, Figure \ref{app:f}, Figure \ref{app:a}), and one MonARCH output (Listing \ref{32out}) is attached in the Appendix. 
	
	When generating the Mandelbrot Set, the main property to note is the time taken to process each section of the image. The top and bottom parts of the image takes less time compared to the middle portion of the image. So each of the partition schemes will require to balance the work load not only by the amount of rows, but also based on the partition scheme that makes use of all the tasks equally.

\begin{table*}
	\centering
	\caption{Analysis of Theoretical Speedup vs Actual Speedup}
	\subfloat[Theoretical Speedup vs Actual Speedup of all the partition schemes]
	{
		\renewcommand{\arraystretch}{1.2}
		\label{ac}
		\begin{tabular}{|p{0.5cm} |p{1.5cm}|p{1.5cm}|p{1.5cm}|p{1.5cm}|}			
			\hline
			\textbf{N} & \multicolumn{1}{p{1.5cm}|}{\textbf{Theoretical Speedup}} & \multicolumn{1}{p{1.5cm}|}{\textbf{Naïve Row-Based}} & \multicolumn{1}{p{1.5cm}|}{\textbf{First Come First Served}} & \multicolumn{1}{p{1.5cm}|}{\textbf{Alternating Row-Based}} \\ \hline
			2  &   1.99440	&1.97812&	1.00396	&1.99000  \\ \hline
			4  &   3.96659	&2.05984	&2.98414	&3.96060  \\ \hline
			8  &    7.84583	&2.49809	&6.77486	&7.60671    \\ \hline
			16 &   15.35350	&3.97305	&13.33375	&13.37449  \\ \hline
			32 &   29.43820	&7.37966	&23.72932	&22.87749 \\ \hline
			$\infty$ &      356.22764   &  - &  - &    -  \\ \hline
			
		\end{tabular}
	}
	\subfloat[Percentage Difference of all the partition schemes]
	{
		\renewcommand{\arraystretch}{1.2}
		\label{pd}
		\begin{tabular}{|p{0.5cm}|p{1.5cm}|p{1.5cm}|p{1.5cm}|}			
			\hline
			\textbf{N} & \multicolumn{1}{p{2cm}|}{\textbf{Naïve Row-Based}} & \multicolumn{1}{p{2cm}|}{\textbf{First Come First Served}} & \multicolumn{1}{p{2cm}|}{\textbf{Alternating Row-Based}} \\ \hline
			2  &   0.81952&	66.06515&	0.22072  \\ \hline
			4  &  63.27981&	28.26920&	0.15114  \\ \hline
			8  &   103.39878&	14.65004&	3.09480   \\ \hline
			16 &  117.77011&	14.08118&	13.77758 \\ \hline
			32 &  119.82522	&21.47507&	25.08121 \\ \hline
			$\infty$ &   - &  - &    -  \\ \hline
			
		\end{tabular}
	}
\end{table*}

	\begin{figure*}[!t]
	\centering
	\caption{Plots for Theoretical Speedup vs Actual Speedup}
	\subfloat[Theoretical Speedup vs Actual Speedup]
	{
		\includegraphics[width=3.3in]{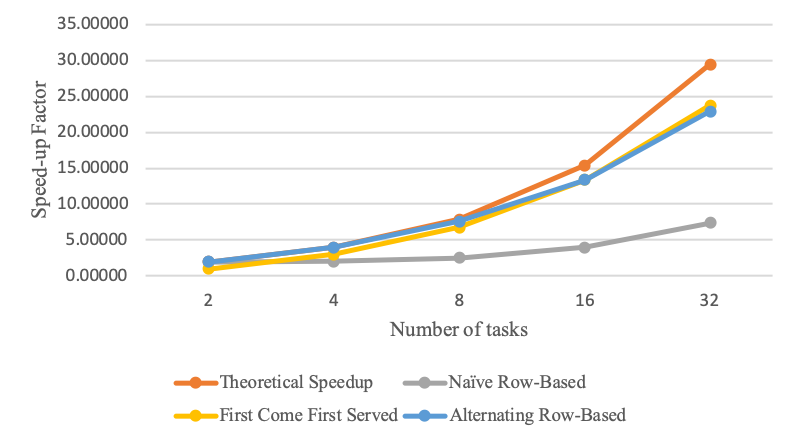}
		
	}
	\subfloat[Percentage Difference]
	{
		
		\includegraphics[width=3.3in]{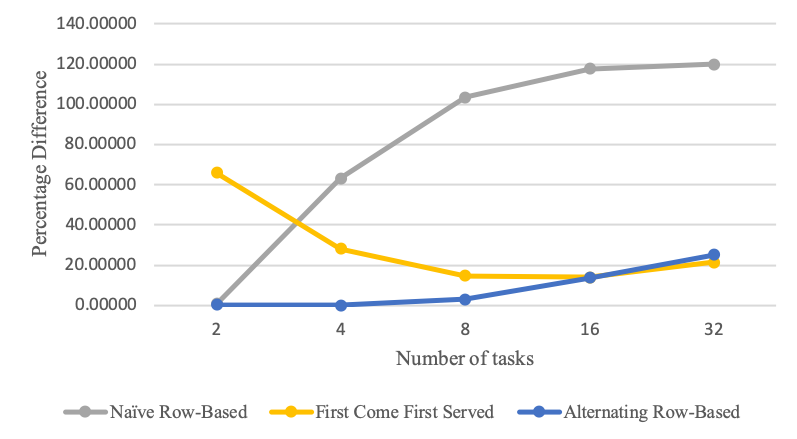}
		\label{dpd}
	}
\end{figure*}

	\section{Results and Discussions}

	Now let us discuss about the results obtained from the testing phase. The actual speed up from Table \ref{ac} is calculated using equation \ref{actualvst}. 
	
	\begin{equation}
		\label{actualvst}
		\textit{Actual Speed up} = \frac{\textit{Total time by single task}}{\textit{Total time by multiple tasks}}
	\end{equation}
	
	Then using the Theoretical Speed up for $n$ tasks as $T_n$  and Actual Speed up as $A_n$ from Table \ref{ac}, we calculate the percentage difference using equation \ref{percen}. 
	
	\begin{equation}
	\label{percen}
	\textit{Percentage Difference} = \frac{|T_n - A_n|}{\frac{T_n + A_n}{2}} \times 100	\end{equation}
	
	Each partition scheme is then compared and discussed based on the following criteria. 
	
	\begin{itemize}
		\item Theoretical Speed up vs Actual Speed up
		\item Actual Speed up comparison between each partition scheme
	\end{itemize}

	\subsection{Theoretical Speed up vs Actual Speed up}
	
	In this section we will compare how the actual speed up compares against the theoretical speed up established in Section \ref{Theo}
	
	\subsubsection{Naive - Row Based Partition Scheme }
	\label{naiveres}
	This method divides the whole image into $N$ number of parts, where $N$ is the number of tasks.
	Referring to Table \ref{ac}, for two tasks, the theoretical speed up and the actual speed up is almost equal. The main reason for this is when we are using two cores, the workload is divided equally among the two tasks. Both the tasks will be doing the high iteration and low iteration part of the image.
	But when the number of tasks increase, the actual speed up becomes poor. The reason for this is, even though the tasks get equal rows, the number of iterations needed for some rows are greater than others, as mentioned previously in Section \ref{TEST}. Due to this, the tasks in-charge of the top and bottom quarter of the image takes less time compared to the middle part, so the master node has to wait until all the tasks are completed to write the file. Finally comparing the percentage difference using Table \ref{pd} we see that this partition scheme does not perform well when the task count increases.
	
	\subsubsection{First Come First Served - Row Based Partition Scheme}
	This method sends the rows based on the availability of the task. Looking back at Table \ref{ac}, we see that the performance of the two tasks test is almost equal to no speed up at all. The main reason for this is when we use two tasks, the slave generates all the rows in the image, while the master sends and receives data. So the actual speedup should be $ \le 1$, but our actual speed up $> 1$. Why? According to a paper published in 2016, by a group of researchers \cite{ristov2016superlinear}, this behavior is common in parallel systems as the serial program needs more RAM to store the data while generating the image; the parallelized system can use the cache as it can store the small chunks of data received at each send and receive. Another explanation for this phenomena is that Amdal's Law does not take into account the communication overhead. Therefore, the speed up can be $\ge 1$ even with a lot of communication.	
	Next, comparing the theoretical speed up vs actual speed up for multiple tasks, we see that this partition scheme is sub-linear. The percentage difference is at a reasonable difference, but when two cores are used, it does not perform well because of the reasons mentioned above. But the percentage difference decreases with each increment of the core count.
	
	\subsubsection{Alternating - Row Based Partition Scheme}
	This method divides the rows based on the rank ($r_n$)  of each task. Referring to Table \ref{ac}, this approach is almost equal to the theoretical speed up when two cores are used. The reason is the same as \ref{naiveres}, where the work load is divided equally among the two tasks. Even when the number of tasks increases, this method is able to perform in a sub-linear order staying close to the theoretical limit. Referring to Table \ref{pd}, we see that the the percentage difference increases with each increment in the number of tasks.
	
	Now that we have an idea about the difference between the Theoretical Speed up and Actual Speed up, let us look into how they compare against each other.
	
	\subsection{Actual Speed up between each Partition Scheme}
	
		\subsubsection{Naïve - Row Based Partition Scheme }
		Comparing the Naïve approach with the rest, we immediately notice the performance is poor. But evaluating dual core performance, this approach performs better than the First Come First Served approach. But as the number of tasks increases, this approach cannot keep up with the other two partition schemes. Therefore, we can conclude that this partition scheme performs the worst out of the three approaches.
	\subsubsection{First Come First Served - Row Based Partition Scheme}
	\label{fcfsex}
		Comparing this approach with the rest, we see that it performs average, but as the number of tasks increases, we see that this performs better than any of the other methods. The reason for this is, both the other methods cause a bottle neck when the master receives the data at the same time from all the slaves. The problem is that, if the work load is divided equally, then all the slaves finish at the same time thus causing a bottle neck at the master node. 
		
		But First Come First Served approach uses more communication to ensure that the bottle neck is amortized over all the communication instead of all at once. So the master node does not have to copy all the rows from each task at the end. It will do it as the tasks finish each row. This leads to the First Come First Served partition approach to be faster when the number of tasks is higher.
	
	\subsubsection{Alternating - Row Based Partition Scheme}
	
	Comparing this approach with the Naïve approach, for two cores we see that the work load distribution is the same. But why is it faster compared to the Naïve approach? The reason is that, in the Naïve approach the master node is computing the high iteration part at the end, so even though the slave finishes at the same time, it has to wait till the master node is ready to receive. In the alternating approach, each task gets the high iterating part in the middle, so they finish at the same time, thus reducing the wait time for the slave.
	
	Referring to Figure \ref{dpd}, we see that the First Come First Served approach, is faster than the Alternating approach when the number of tasks is increased. The reason for this is explained in \ref{fcfsex}. We see that more communication leads to slower speed up for the First Come First Served approach but as the number of tasks increased, both Naive and Alternating methods started slowing down. This can also be due to the increase in the total number of communication between the master and slave for the Naïve and Alternating approaches, whereas it was constant for the First Come First Served approach. This was represented by equations \ref{naive_m}, \ref{fcfs_m} and \ref{al_m}.

	\section{Conclusion}
	Now that we have analyzed all the cases, we can rank each of the partition schemes based on their performance. 
	
	Referring to Figure \ref{dpd}, the Naïve Row Based Segmentation performed the worst with increasing percentage difference at each doubling of the number of tasks except when the number of tasks were two. Next, the First Come First Served partition scheme performed on average for most of the test cases but performed the best when the number of tasks were $ > 16$. Finally, the Alternating Row Based partition scheme performed the best when the number of tasks $ > 16$. 
	
	The speed up differences between each partition scheme can be explained due to the bottle necks and communication overhead of each approach. The partition scheme that is able to reduce each of the factors is able to perform better than the rest.
	
	Conclusively, we can state that First Come First Served partition scheme is effective when the number of tasks are high while the Alternative partition scheme is effective when the number of tasks are low.
	
	\section{Future Work}
	This report opens up future research paths for each of the three methods. The effect of having the number of tasks greater than 32 can be a one of areas that can be experimented in the future. Another path is removing the assumption of having tasks as multiples of two and having an odd number of tasks. This will allow us to observe how the remainder affects each of the methods, especially Naïve and Alternative since the workload will be imbalanced for some nodes. 
	This reports paves the way to exploring the efficient ways of generating the Mandelbrot Set in parallel architectures.

\bibliographystyle{IEEEtran}
\bibliography{Report.bib}

	\newpage
	\onecolumn
	\appendix
		\begin{figure*}[h]
		\centering
		\includegraphics[width=7cm,keepaspectratio]{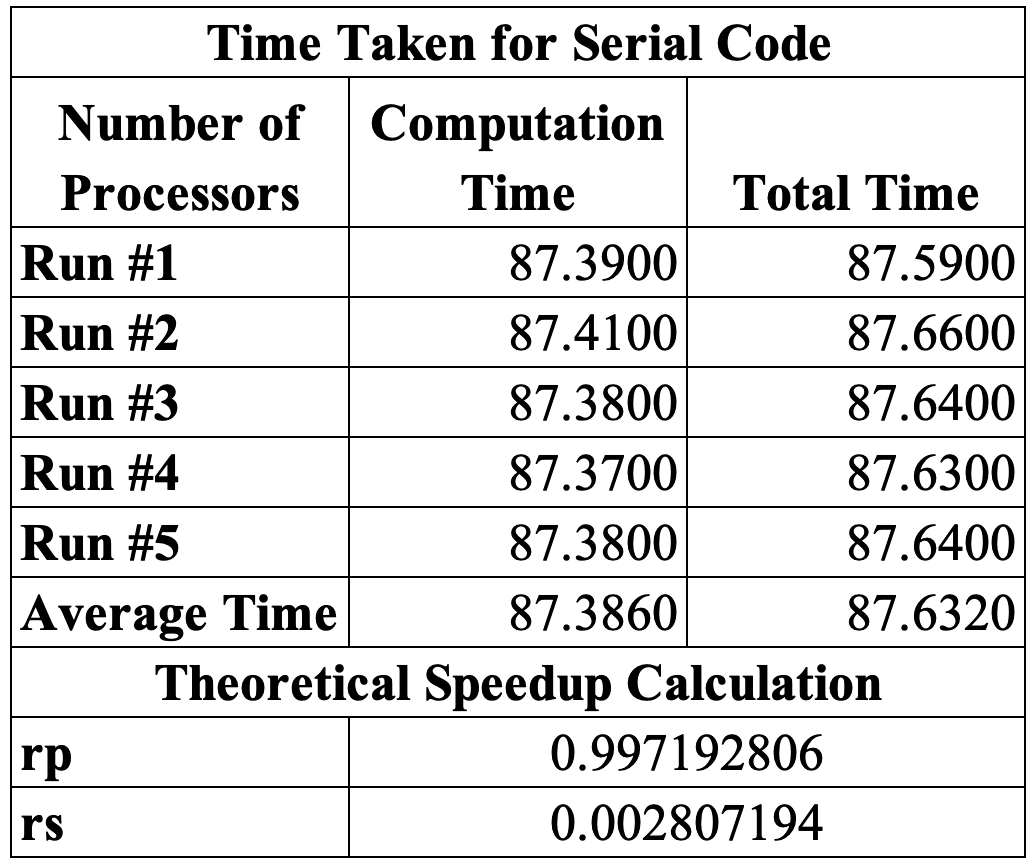}
		\caption{Test Results : Computation vs Total Time for Serial Generation}
		\label{app:amdal}
		\caption*{\emph{**All times are expressed in seconds}}
	\end{figure*}
\begin{lstlisting}[language={}, caption={Sample MonARCH output file for 32 cores},label={32out}]
Alternative : 32 cores
Run 1
File: Mandelbrot.ppm successfully opened for writing.
Computing Mandelbrot Set. Please wait...
Mandelbrot computational process time: 2.855784
Completed Computing Mandelbrot Set.
File: Mandelbrot.ppm successfully closed.
Mandelbrot total process time: 4.238303
Run 2
File: Mandelbrot.ppm successfully opened for writing.
Computing Mandelbrot Set. Please wait...
Mandelbrot computational process time: 2.857663
Completed Computing Mandelbrot Set.
File: Mandelbrot.ppm successfully closed.
Mandelbrot total process time: 3.709431
Run 3
File: Mandelbrot.ppm successfully opened for writing.
Computing Mandelbrot Set. Please wait...
Mandelbrot computational process time: 2.858120
Completed Computing Mandelbrot Set.
File: Mandelbrot.ppm successfully closed.
Mandelbrot total process time: 3.726426
Run 4
File: Mandelbrot.ppm successfully opened for writing.
Computing Mandelbrot Set. Please wait...
Mandelbrot computational process time: 2.855122
Completed Computing Mandelbrot Set.
File: Mandelbrot.ppm successfully closed.
Mandelbrot total process time: 3.742856
Run 5
File: Mandelbrot.ppm successfully opened for writing.
Computing Mandelbrot Set. Please wait...
Mandelbrot computational process time: 2.855696
Completed Computing Mandelbrot Set.
File: Mandelbrot.ppm successfully closed.
Mandelbrot total process time: 3.735431
\end{lstlisting}

\newpage
\begin{lstlisting}[caption={MonARCH job script for 16 cores},label={16job}]
#!/bin/bash
#SBATCH --job-name=alter_16
#SBATCH --time=00:30:00
#SBATCH --mem=32G
#SBATCH --ntasks=16
#SBATCH --cpus-per-task=1
#SBATCH --ntasks-per-node=16
#SBATCH --account=fit3143
#SBATCH --constraint=Xeon-Gold-6150
#SBATCH --output=alter_16.out
module load openmpi/1.10.7-mlx

echo "Alternative : 16 cores"

echo "Run 1"
srun mandelbrot_parallel_alternating

echo "Run 2"
srun mandelbrot_parallel_alternating

echo "Run 3"
srun mandelbrot_parallel_alternating

echo "Run 4"
srun mandelbrot_parallel_alternating

echo "Run 5"
srun mandelbrot_parallel_alternating

\end{lstlisting}

\begin{lstlisting}[caption={MonARCH job script for 32 cores},label={32job}]
#!/bin/bash
#!/bin/bash
#SBATCH --job-name=alter_32
#SBATCH --time=00:30:00
#SBATCH --mem=32G
#SBATCH --ntasks=32
#SBATCH --cpus-per-task=2
#SBATCH --ntasks-per-node=16
#SBATCH --account=fit3143
#SBATCH --constraint=Xeon-Gold-6150
#SBATCH --output=alter_32.out
module load openmpi/1.10.7-mlx

echo "Alternative : 32 cores"

echo "Run 1"
srun mandelbrot_parallel_alternating

echo "Run 2"
srun mandelbrot_parallel_alternating

echo "Run 3"
srun mandelbrot_parallel_alternating

echo "Run 4"
srun mandelbrot_parallel_alternating

echo "Run 5"
srun mandelbrot_parallel_alternating



\end{lstlisting}
\newpage
	\begin{sidewaysfigure*}[h]
	\centering
	\includegraphics[width=\textheight,height=\textwidth,keepaspectratio]{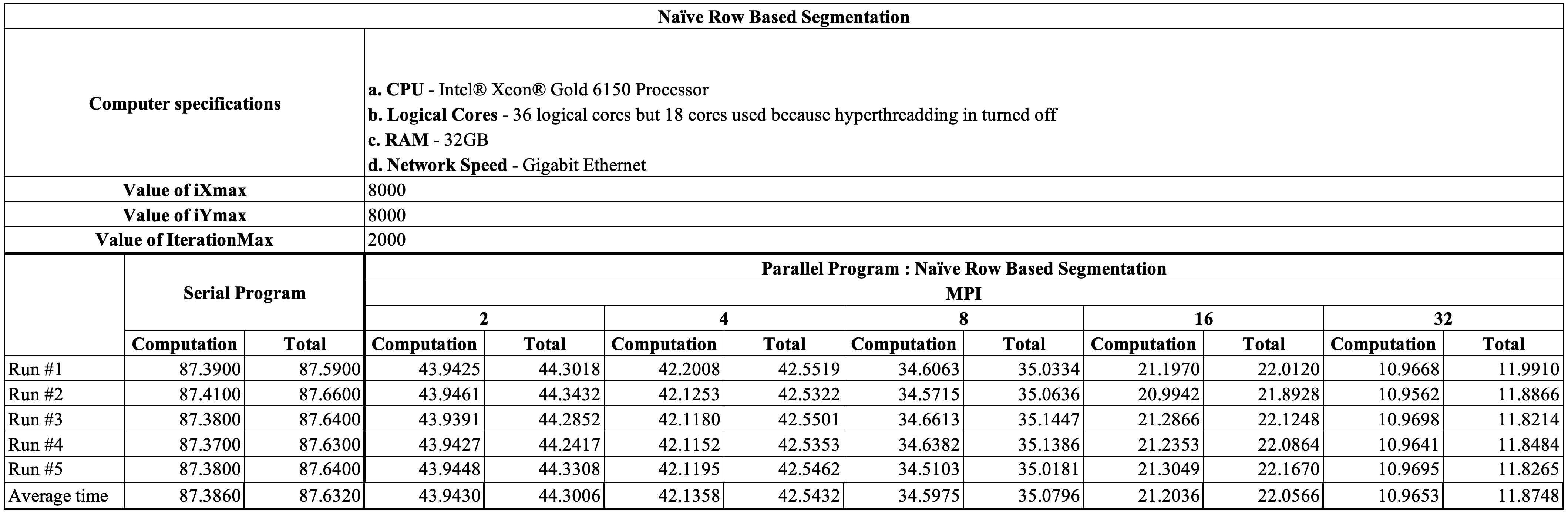}
	\caption{Test Results : Naïve Row Based Segmentation}
	\label{app:n}
	\caption*{\emph{**All times are expressed in seconds}}
\end{sidewaysfigure*}
\newpage
\begin{sidewaysfigure*}[ht]
	\centering
	\includegraphics[width=\textheight,height=\textwidth,keepaspectratio]{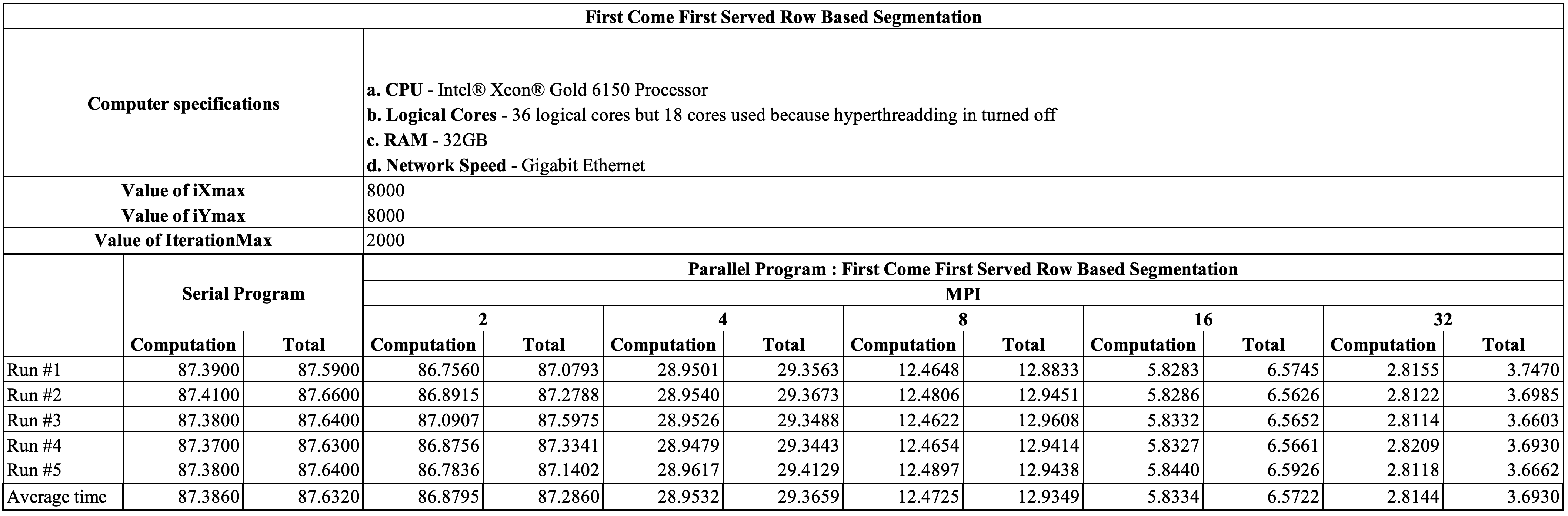}
	\caption{Test Results : First Come First Served Row Based Segmentation}
	\label{app:f}
	\caption*{\emph{**All times are expressed in seconds}}
\end{sidewaysfigure*}
\newpage
\begin{sidewaysfigure*}[ht]
	\centering
	\includegraphics[width=\textheight,height=\textwidth,keepaspectratio]{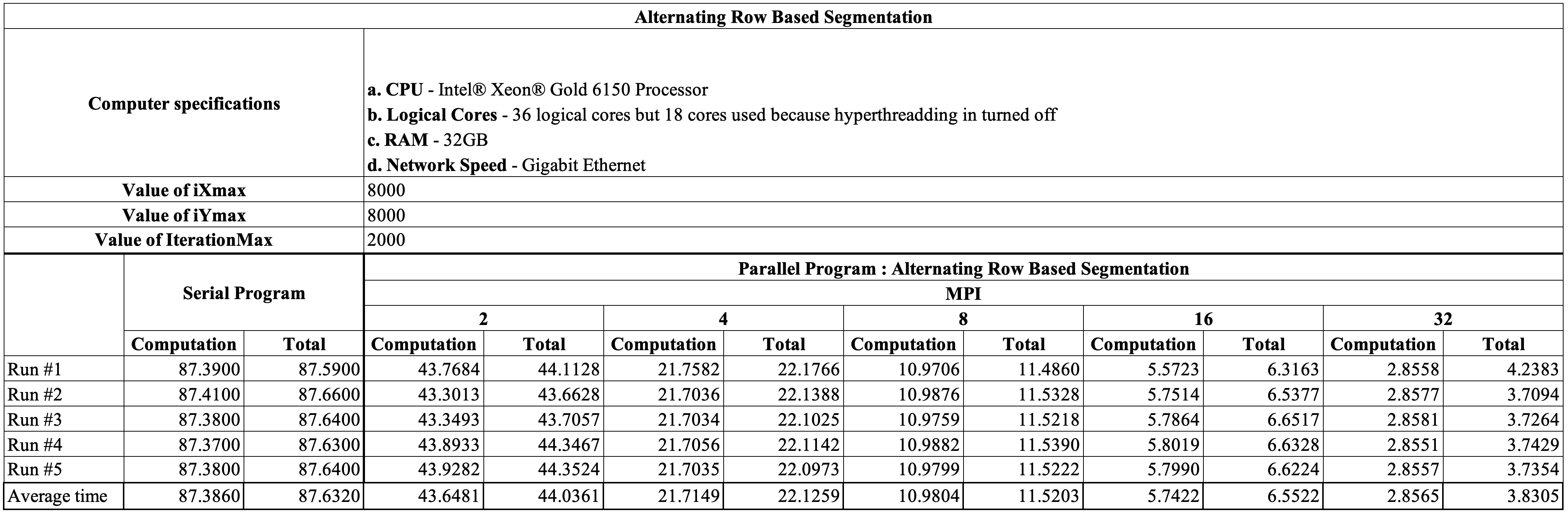}
	\caption{Test Results : Alternating Row Based Segmentation.}
	\label{app:a}
	\caption*{\emph{**All times are expressed in seconds}}
\end{sidewaysfigure*}

\end{document}